\documentclass[aps,prl,twocolumn,preprintnumbers,superscriptaddress,amsmath]{revtex4}
\usepackage{amssymb}
\usepackage{txfonts}
\usepackage{amssymb}
\usepackage{graphicx}
\usepackage{subfigure}
\usepackage{dcolumn}
\usepackage{bm}

\begin{document}

\title{Generation of multiphoton entangled quantum states with a single silicon nanowire}

\author{Ming Zhang\footnote[1]}
\affiliation{State Key Laboratory for Modern Optical Instrumentation, Centre for Optical and Electromagnetic Research,
Zhejiang Provincial Key Laboratory for Sensing Technologies, Zhejiang University, Zijingang Campus, Hangzhou
310058, China.}
\author{Lan-Tian Feng\footnote[1]{These authors contributed equally to this work.}}
\author{Zhi-Yuan Zhou}
\author{Yang Chen}
\affiliation
{Key Laboratory of Quantum Information, University of Science and Technology of China, CAS, Hefei, 230026, People's Republic of China.}
\affiliation{Synergetic Innovation Center of Quantum Information $\&$ Quantum
Physics, University of Science and Technology of China, Hefei, Anhui
230026, China.}
\author{Hao Wu}
\author{Ming Li}
\affiliation{State Key Laboratory for Modern Optical Instrumentation, Centre for Optical and Electromagnetic Research,
Zhejiang Provincial Key Laboratory for Sensing Technologies, Zhejiang University, Zijingang Campus, Hangzhou
310058, China.}
\author{Guo-Ping Guo}
\author{Guang-Can Guo}
\affiliation
{Key Laboratory of Quantum Information, University of Science and Technology of China, CAS, Hefei, 230026, People's Republic of China.}
\affiliation{Synergetic Innovation Center of Quantum Information $\&$ Quantum
Physics, University of Science and Technology of China, Hefei, Anhui
230026, China.}
\author{Dao-Xin Dai\footnote[3]{dxdai@zju.edu.cn}}
\affiliation{State Key Laboratory for Modern Optical Instrumentation, Centre for Optical and Electromagnetic Research,
Zhejiang Provincial Key Laboratory for Sensing Technologies, Zhejiang University, Zijingang Campus, Hangzhou
310058, China.}
\author{Xi-Feng Ren\footnote[2]{renxf@ustc.edu.cn}}
\affiliation
{Key Laboratory of Quantum Information, University of Science and Technology of China, CAS, Hefei, 230026, People's Republic of China.}
\affiliation{Synergetic Innovation Center of Quantum Information $\&$ Quantum
Physics, University of Science and Technology of China, Hefei, Anhui
230026, China.}

\maketitle
\textbf{Multiphoton entanglement plays a critical role in quantum information processing, and greatly improves our fundamental understanding of the quantum world. Despite tremendous efforts in either bulk media or fiber-based devices, nonlinear interactions in integrated circuits show great promise as an excellent platform for photon pair generation with its high brightness, stability and scalability \cite{Caspani2017}. Here, we demonstrate the generation of bi- and multiphoton polarization entangled qubits in a single silicon nanowire waveguide, and these qubits directly compatible with the dense wavelength division multiplexing in telecommunication system. Multiphoton interference and quantum state tomography were used to characterize the quality of the entangled states. Four-photon entanglement states among two frequency channels were ascertained with a fidelity of $0.78\pm0.02$. Our work realizes the integrated multiphoton source in a relatively simple pattern and paves a way for the revolution of multiphoton quantum science.}

Optical quantum states with entanglement shared among several modes are critical resources for studies in quantum communication \cite{Kimble2008}, computation \cite{Walther2005,Humphreys2013}, simulation \cite{Aspuru-Guzik2012}, and metrology \cite{Afek2010}. Therefore, the controllable and scalable realization of multiphoton quantum states would enable a practical and powerful implementation of quantum technologies. For generating  and manipulating photon pairs, nonlinear interactions in integrated circuits show great promise as an excellent platform with its high brightness, stability and scalability. Recently, optical integrated kerr frequency combs in high-refractive-index glass platform were used to generate time-bin bi- and multiphoton entangled qubits \cite{Reimer2016}, which started a new time to exploit multiphoton entangled states with the help of quantum photonic integrated circuits.

The silicon-on-insulator (SOI) photonic circuit is a very promising platform for realizing complex quantum states for its strong third-order optical nonlinearity, low nonlinear noise, small footprint, mature fabrication technics and compatibility with complementary metal oxide semiconductor (CMOS) electronics \cite{Bogaerts2005} as well as telecom techniques. Multiple biphoton entanglement sources have been manipulated and further used for large-scale integration and quantum information processings \cite{Li2017,Silverstone2014,Harris2014,Wang2016,Wang2017,Paesani2017}. Nevertheless, whether the silicon photonic circuit can be used to manipulate multiphoton entangled source is still unexplored.

\begin{figure*}[htbp]
\centering
\includegraphics[width=14.0cm]{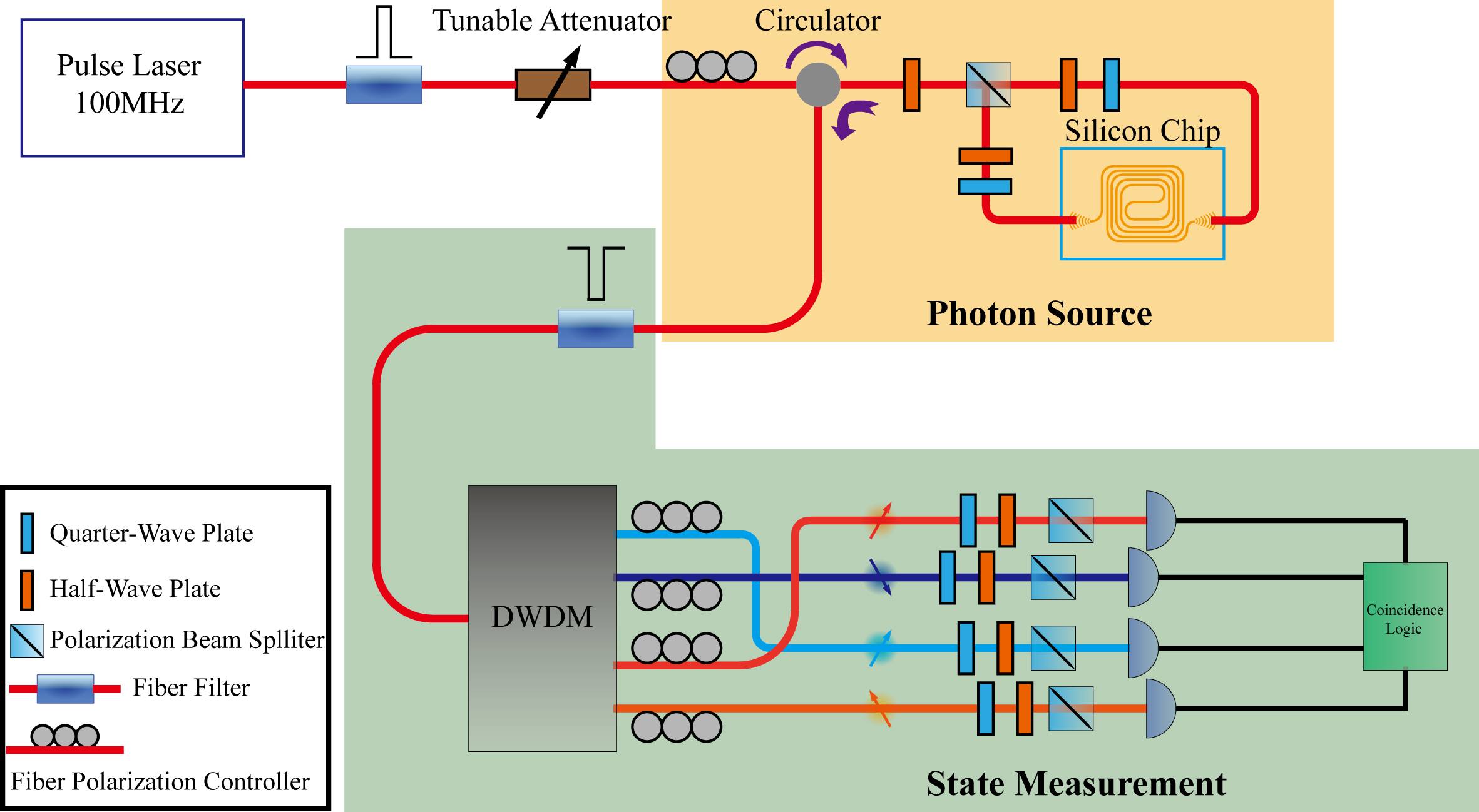}
 \caption {\textbf{Schematic showing the generation and measurement of the multiphoton entanglement states from a silicon nanowire.} A pulse erbium laser with a repetition frequency $100 MHz$ was involved as the pump light. After forward and backward filtering with $100 GHz$ bandwidth, it was input into a Sagnac loop for the generation of polarization entanglement states. One DWDM system was used to divide photon pairs into multiple frequency channels, and normal architectures for polarization state tomography were used to ascertain the quality of the entangled quantum states.}
\end{figure*}

In this study, we use the silicon-on-insulator (SOI) photonics platform to explore the generation of multiphoton entangled states. A single silicon nanowire waveguide, which has a length of $1\ cm$ and transverse dimension of $\sim450\ nm\times220\ nm$, is employed to generate photon pairs. The external pump laser is coupled to the waveguide with grating couplers. The waveguide has a propagation loss of $1\ dB$, and the total insertion loss of this structure is $11\ dB$, including $5\ dB$ coupling loss each at the input and output grating coupler.

Polarization encoding was chosed to demonstrate the multiphoton entanglement generation, since it is popular for various quantum information applications \cite{Kok2007}. The experimental setup for generation and measurement of the multiphoton polarization entangled states is shown in Fig. 1. A pulsed erbium laser with a repetition frequency $100\ MHz$ was involved for the degenerate spontaneous four wave mixing process in the silicon nanowire. After forward and backward filtering with $100\ GHz$ bandwidth, the pump light was passed through a polarization controller (PC) and a fiber circulator, and then went into a Sagnac loop \cite{Takesue2008}. The Sagnac loop consists of half-wave plates (HWPs), quarter-wave plates (QWPs), a polarization beam splitter (PBS) and the chip of silicon nanowire. The combination of the HWP and QWP is used to modulate the optical polarization for maximum grating coupling. In the Sagnac loop, pump beam was split into clockwise and counterclockwise circulation directions by the PBS, and the HWP before the PBS is used for optical power modulation. In each direction, time correlated signal and idler photons are generated and their frequencies are equally separated from the central pump frequency. In the clockwise circulation direction, the photon pair generated is $\left|V_sV_i\right\rangle$, while in the counterclockwise circulation direction, the photon pair generated is $\left|H_sH_i\right\rangle$ after the PBS. The biphoton quantum state at the output port of the loop can be expressed as
\begin{equation}\label{1}
  \left|\Phi\right\rangle=\frac{1}{\sqrt{1+\eta^2}}(\left|H_sH_i\right\rangle+\eta e^{i\delta}\left|V_sV_i\right\rangle),
\end{equation}
where $\eta$ is determined by the ratio of the pump power in the two circulation directions, and $\delta$ depends on the initial phase of the pump beam and the birefringence experienced by the pump beam in the $H$ and $V$ polarizations. By modulating the pump power ratio and relative phase, a maximally polarization entangled Bell state
\begin{equation}\label{2}
  \left|\Phi\right\rangle=\frac{1}{\sqrt{2}}(\left|H_sH_i\right\rangle+\left|V_sV_i\right\rangle)
\end{equation}
can be produced.

At the output port of the fiber circulator, the pump laser was blocked by post-filters with $200\ GHz$ bandwidth. One 40-channel dense wavelength division multiplexing (DWDM) system was used to separate the signal and idler photons, whose frequencies are equally separated from the central pump frequency.. Each channel has a $100\ GHz$ bandwidth, then we can select any combination of photon pairs with a frequency detuning about $2\ THz$ from the central pump frequency. The generated photon pairs were finally detected by superconducting nanowire single-photon detectors (SCONTEL, dark count rate $100\ Hz$, detector efficiency $85\%$ at C band).

The entangled photon pairs are frequency multiplexed and generated over all DWDM frequency channels \cite{Li2017}. We selected five signal-idler channels to ascertain the effectiveness and stability of our system (Supplementary Table 1). Through inputting $120\ \mu W$ pump light into the single cycle of the Sagnac loop, we recorded two-photon coincidences between different signal-idler channel combinations (Supplementary Fig. 1). It shows that the crosstalk is negligible between different frequency channels, even for the adjacent ones.

Next, we modulated the polarization of the pump light before the PBS of the Sagnac loop to obtain the maximally entangled states (Eq. 2). Combinations of QWP, HWP and PBS after the DWDM were used to constitute a normal polarization state tomography architecture \cite{James2001} for characterizing the degree of entanglement. We took the signal-idler channels 1 for testing the quality of the entanglement. We first measured the two-photon polarization interference fringes. The interference is expected to be proportional to $1+Vsin(2\pi(x-x_c)/T)$, where $V$ is the fringe visibility, $x_c$ is the initial phase, and $T$ is the oscillation period. The fringe visibility $V$ is defined as $V=(d_{max}-d_{min})/(d_{max}+d_{min})$, where $d_{max}$, $d_{min}$ are the maximum and minimum fitting data, respectively. Through setting the angle of the HWP in the signal channel as $0^\circ$ or $45^\circ$, and modulating the angle of the HWP in the idler channel, we obtained two interference fringes (Fig. 2a). The raw visibilities in the $0^\circ$ (solid red line) and $45^\circ$ (solid black line) bases were $96.1\pm3.2\%$ and $93.0\pm3.2\%$, respectively, which, being greater than $\frac{1}{\sqrt{2}}\approx70.7\%$, confirming entanglement through the violation of the Clauser-Horne-Shimony-Holt (CHSH) inequality. In this case, we estimated the photon pair generation rate of $270\ kHz$ per channel (0.0027 pairs per double pulse), accounting for system and detection losses of $16\ dB$ (see methods).

\begin{figure}[htbp]
\centering
\includegraphics[width=8.0cm]{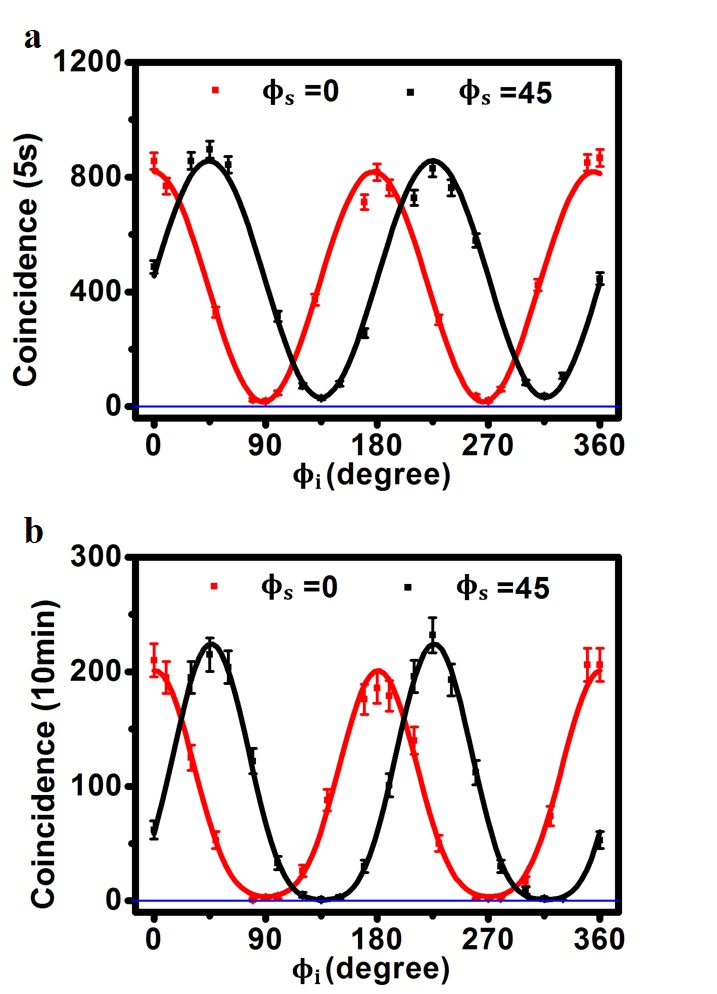}
\caption {\textbf{Quantum interference fringes of two-photon and four-photon polarization entangled states.} Two-fold and four-fold coincidences as a function of the idler polarizer angles when the signal polarizer angles were kept at $0^\circ$ (solid red line) and $45^\circ$ (solid black line), respectively. Four-photon entangled states unfold totally different interference patterns from the two-photon entangled states. The error bar was obtained from the square root of the experiment data.}
\end{figure}

Now we consider the multiphoton cases. In principle, the generated multiphoton entanglement state for our experimental setup can be expressed as
\begin{equation}\label{3}
  \left|\Phi^{2n}\right\rangle=\frac{1}{\sqrt{2^n}}(\left|H_{s1}H_{i1}\right\rangle+\left|V_{s1}V_{i1}\right\rangle)\otimes\cdots\otimes(\left|H_{sn}H_{in}\right\rangle+\left|V_{sn}V_{in}\right\rangle).
\end{equation}
That is, combination of different correlated signal-idler channels could be used to produce multiphoton entangled states. For example, by selecting the signal-idler channels 1 and 5, we can get two two-photon entangled states, given by $\left|\Phi_1\right\rangle=\frac{1}{\sqrt{2}}(\left|H_{s1}H_{i1}\right\rangle+\left|V_{s1}V_{i1}\right\rangle)$ and $\left|\Phi_5\right\rangle=\frac{1}{\sqrt{2}}(\left|H_{s5}H_{i5}\right\rangle+\left|V_{s5}V_{i5}\right\rangle)$, respectively. By post-selecting four-photon events with one photon on each frequency channel, these two states are multiplied, resulting in four-photon polarization entangled states, which is given by
\begin{equation}\label{4}
  \left|\Phi^{4}\right\rangle=\frac{1}{2}(\left|H_{s1}H_{i1}\right\rangle+\left|V_{s1}V_{i1}\right\rangle)\otimes(\left|H_{s5}H_{i5}\right\rangle+\left|V_{s5}V_{i5}\right\rangle).
\end{equation}
To prove this state, we measured the four-photon quantum interference fringes, which generally cannot be present for two completely independent two-photon qubit states. By setting the pump power to $600\ \mu W$, we obtained a four-fold coincidence rate of $0.34\ Hz$, which corresponds to a calculated generation rate of $340\ kHz$, taking into account the system and detection losses of $15\ dB$ (see methods).  The interference is expected to be proportional to $1+(1-\sqrt{1-V^2})/V)sin^2(2\pi(x-x_c)/T$. By setting the angle of the HWPs in the signal channels as $0^\circ$ or $45^\circ$, and modulating the angle of the HWPs in the idler channels simultaneously, we obtained two four-photon interference fringes (Fig. 2b). The raw visibilities in the $0^\circ$ (solid red line) and $45^\circ$ (solid black line) bases were $96.5\pm1.5\%$ and $99.1\pm0.4\%$, respectively. Four-photon entangled states unfold totally different interference patterns from the two-photon entangled states and the high interference visibilities prove the existence of the multiphoton entanglement. Note that, for the measurement of quantum interference fringes, the QWPs were moved out from the setup.

\begin{figure*}[htbp]
\centering
\includegraphics[width=17.0cm]{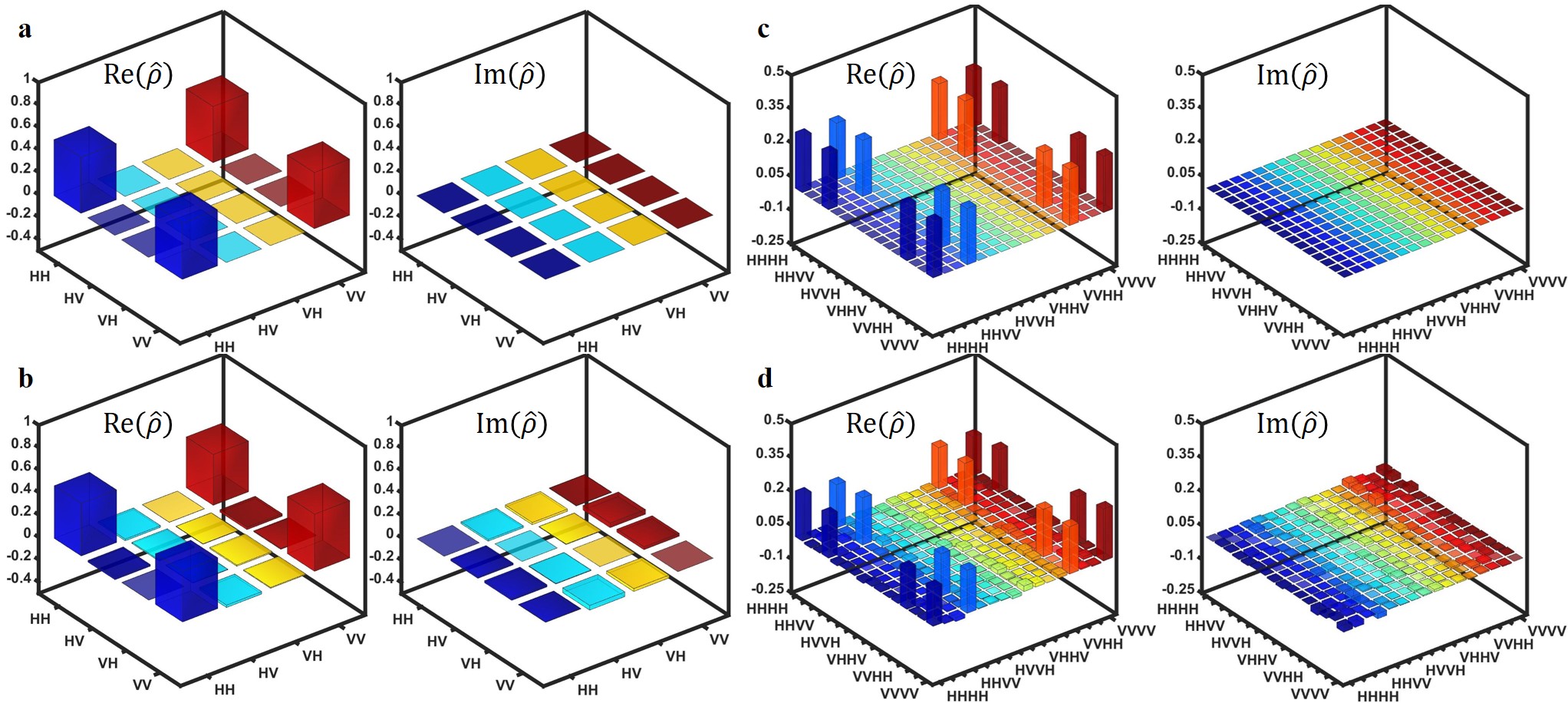}
\caption {\textbf{Quantum state characterization through state tomography.} A quantum state can be fully described by its density matrix $\widehat{\rho}$. The real (Re) and imaginary
(Im) parts of the ideal density matrices of two- and four-photon entangled qubit states are shown in \textbf{a} and \textbf{c}, respectively. The measured density matrix of the two-photon states \textbf{b} agrees well with the ideal case. The measured density matrix of the four-photon states \textbf{d} reaches a fidelity of $0.78\pm0.02$, which is completely satisfied for quantum information processings.}
\end{figure*}

Quantum state tomography \cite{James2001} was also used to fully characterize the entangled states. Using this method, we could obtain the experimental state density matrix and thus study how close of the measured states to the ideal entangled states. Firstly, we performed the two-photon quantum state tomography for the frequency channels 1. 16 data at different measurement bases were acquired to construct the state density matrix. The ideal density matrix of the maximally entangled states (Eq. 2) and the measured density matrix of the output states were displayed in Figs. 3a, and 3b, respectively. We used the maximum-likelihood-estimation method to construct the density matrix with the experimental results. The fidelity is defined as $F=Tr(\widehat{\rho}_{exp}\widehat{\rho}_{th})$, where $Tr$ is the trace, $\widehat{\rho}_{exp}$ is the measured density matrix, and $\widehat{\rho}_{th}$ is the ideal density matrix. We estimated the fidelity of $0.94\pm0.01$, confirming that the generated quantum states were of high quality and very close to the ideal maximally entangled states. The error of the fidelity was obtained by 100 times Monto Carlo calculation with the experimental data subject to Gaussian statistics. The deviation of the fidelity from unity was mainly due to the unideal rotating the angle of the wave plates. It is worth pointing out that the quantum states are frequency multiplexed and entanglement is in any signal-idler channels. As a test, we also performed the two-photon quantum state tomography for signal-idler channels 3 and 5, and found the fidelities of $0.97\pm0.01$ and $0.95\pm0.01$, respectively (Supplementary Fig. 1).

For the four-photon entangled states, 256 data of different measurement bases were obtained to construct the state density matrix. The ideal density matrix of the maximally entangled states (Eq. 4) and the measured density matrix of the output states from signal-idler channels 1 and channel 5 were displayed in Figs. 4c, and 4d, respectively. We obtained a fidelity of $0.78\pm0.02$ without background and accidence subtraction. The high visibility clearly proved the entanglement and is satisfied for further practical applications.

Since it was demonstrated recently that the frequency encoding can be used for single-photon frequency shifting \cite{Fan2016} and construction of high-fidelity quantum gates  \cite{Lukens2017,Lu2018} for high-dimensional quantum information processing, our  presented multiphoton entangled source is fully compatible with these post-processing systems and it will constitutes an indispensable block for the frequency-based quantum information processing. Furthermore, the multiphoton entangled state (Eq. 3) can be converted to the Greenberger-Horne-Zeilinger (GHZ) state through post operation and selection, which is usually used as the resource for linear-optical quantum computation \cite{WangX2016}.

In addition, the entangled photon numbers can be increased \cite{Reimer2016}. Further integration of multiphoton manipulation block in one integrated circuits \cite{Matthews2009} will also lead to more compact and stable systems with higher performance, resulting better detection rates and higher fidelity. A scheme to decrease coupling loss by using high coupling efficiency etched facet tapers \cite{Cardenas2014} and low-loss filter system would increase the brightness of four-fold coincidence greatly, even to the useful kilohertz range.

In conclusion, we have experimentally shown that a single silicon nanowire can be used for multiphoton source manipulation. Due to its strong third-order nonlinearity, we achieved a high brightness photon pair source with a very low pump power. The fidelity of the four-photon entangled states is high enough for practical applications. The multiphoton entangled source is directly compatible with the dense wavelength division multiplexing communication system and frequency-based post-processing system, thus provided a scalable and practical platform for quantum information processing.

\section*{Methods}

\noindent {\bf System efficiency.} We ascertained the efficiencies of all departments with laser light measurements. The grating coupler has coupling loss of $5\ dB$. The system for state manipulation and state measurement has loss of $4.3\ dB$. The post-filters and WDMs have inherent loss of $6\ dB$ and both detectors have efficiency of $85\%\ (-0.7\ dB)$. The total loss is $16\ dB$ for two-photon state measurements and $15\ dB$ for four-photon state measurements where we decreased the loss of $1\ dB$ in post-processing system through finer adjustment.

\noindent {\bf Optical apparatus.} We used pulsed erbium laser (repetition frequency $100\ MHz$) to generate the pump light (Supplementary Fig. 2). After filtered, the left laser was amplified by an erbium-doped fiber amplifier and filtered again before input into the source manipulation setup. Fiber alignment was maintained using a piezo-controlled four-dimensional displacement table for position and coupling angle adjustment. The coupling angle was set as $10^\circ$ for fiber-chip coupling. Two cascaded off-chip post-filters ($100\ dB$ extinction ratio) were used to remove the pump photons, and one DWDM system ($30\ dB$ extinction ratio for adjacent channel and $50\ dB$ for non-adjacent channel) was used to separate the signal and idler photons. The correlated photons were recorded by superconducting nanowire single-photon detectors. The electrical signals were collected and analyzed through time-correlated single photon counting (TCSPC) system, and the coincidence window was set as $0.8\ ns$ for two-photon state measurements. For four-photon state measurements, the electrical signals were analyzed by the UQD-Logic and the coincidence window was set as $1\ ns$.

\section*{Acknowledgments} This work was supported by the National Natural Science Foundation of China (NSFC) (Nos. 61590932, 11774333, 61725503, 61431166001), Anhui Initiative in Quantum Information Technologies (No. AHY130300), the Strategic Priority Research Program of the Chinese Academy of Sciences (No. XDB24030600), the National Key R \& D Program (No. 2016YFA0301700),  Zhejiang Provincial Natural Science Foundation of China (Z18F050002), and the Fundamental Research Funds for the Central Universities. This work was partially carried out at the USTC Center for Micro and Nanoscale Research and Fabrication.

\section*{Author contributions}

All authors contributed extensively to the work presented in this paper. M.Z., Z.Y.Z., H.W., M.L. and D.X.D. prepared the samples, L.T.F., M.Z., D.X.D. and X.F.R. performed the measurements, data analyses and discussions. Y.C., G.P.G. and G.C.G. conducted theoretical analysis. X.F.R. and
D.X.D. wrote the manuscript and supervised the project.

\section*{Additional information}

Supplementary information is available in the online version of the paper. Correspondence and requests for materials should be addressed to D.X.D or X.F.R.

\section*{Competing financial interests}
The authors declare no competing financial interests.

\end{document}